# Tunable topologically protected waveguiding in auxetic nonlinear metamaterials

M. Morvaridi[1], F. Bosia[1*], M. Brun[2*], V.F. Dal Poggetto[3], A.S. Gliozzi[1], M. Miniaci[4], C. Croënne[4], N. M. Pugno[3,5], G. Carta[2]

[1] Department of Applied Science and Technology, Politecnico di Torino, Corso Duca degli Abruzzi 24, 10129 Torino, Italy

[2] Department of Mechanical, Chemical and Materials Engineering, University of Cagliari, Piazza d'Armi, 09123 Cagliari, Italy

[3] Laboratory for Bioinspired, Bionic, Nano, Meta Materials and Mechanics, Department of Civil, Environmental and Mechanical Engineering, University of Trento, Trento, 38123, Italy

[4] CNRS, Univ. Lille, Ecole Centrale, ISEN, Univ. Valenciennes, IEMN - UMR 8520, Lille, France

[5] School of Engineering and Materials Science, Queen Mary University of London, London, E1 4NS, UK

* Corresponding authors: federico.bosia@polito.it, mbrun@unica.it

**Abstract**:

In this paper, we discuss the possibility of achieving tunable topologically protected edge modes through the application of uniaxial deformation in an auxetic metamaterial. The proposed structure consists of a thin slab with oriented cuts in a hexagonal lattice, where topologically protected band gaps are opened by introducing a controlled variation in the cut lengths. Numerical simulations demonstrate the existence of topologically protected and scatter-free wave propagation in the structure at the interface between two sub-domains with modified cells, in distinct frequency ranges. This only happens in the presence of auxeticity. In addition, exploiting geometrical nonlinearity, the application of a uniaxial strain can be used to close the topological band gaps or to modify their frequency range, i.e., to weaken the localization effects or to shift the frequency at which they occur. The spatial and temporal variation of the applied strain field can thus be used for the dynamic tuning of metamaterial topological waveguiding properties, with applications in mechanical devices for logic operations and computations.

1. Introduction

Auxetic materials (i.e., materials with negative Poisson's ratio) exhibit the interesting property of expanding/contracting laterally when they are stretched/compressed longitudinally [1–4]. This property leads to an unconventional mechanical behaviour that can be exploited in various fields such as biomedical engineering [5–7], energy harvesting [8–11], textiles and sporting goods [12], armours and ballistic protection [13]. More recently, so-called auxetic metamaterials [14,15] have been studied for their enhanced mechanical properties compared to conventional materials, like indentation resistance [16], impact energy absorption [17,18], and fatigue performance [19,20]. Recent studies have shown that the construction of a hierarchical structure can lead to enhanced auxetic properties in quasi-static metamaterials [21]. In previous work, we have shown both numerically and experimentally that a porous structure with hierarchical cuts can provide extremely negative Poisson's ratio values by activating kirigami-like behaviour and exploiting the interaction between scales, thus strengthening the potential for the mentioned applications [22].

Up to now, most of the investigations on the properties of auxetic materials and metamaterials have focused on their static behaviour, but there are indications that many attractive features also appear in their dynamic properties [23–25], including those in the field of smart transformation optics [26]. Auxetic metamaterials are thus ideal candidates to explore new possibilities in the control of elastic waves, allowing to combine attractive quasi-static and dynamic properties.

In this context, topological protection has recently emerged as a unique means to propagate waves through sharp corners or bends [27,28], being robust with respect to structural disorder [29,30], line defects [31], or backscattering. Initially introduced in the field of condensed matter physics [32,33], topological protection has then been largely investigated in classical systems, such as photonics [34], mechanics [35] and elastic waves [36]. Topological elastic metamaterials are analogous to their electronic and optical counterparts, including Quantum Hall [37–39], Quantum Spin Hall [40,41], and Quantum Valley Hall [42] effects, supporting chiral [43–45], helical [41,46], and valley modes [47–49], respectively. Topologically protected modes arise from the breaking of specific classes of symmetry in correspondence of a Dirac cone [50] and have been exploited for the design of lossless waveguides in phononics [46,51] and mechanical metamaterials [50,52–55].

Despite the high potential demonstrated so far by topological metamaterials, most of the proposed approaches lack dynamic tunability, meaning that their operational frequency cannot be changed once the device is designed. To address this issue, tunable metamaterials [56], whose wave manipulating properties can be modified by various types of actuation (mechanical stimuli, heat transfer, chemical reaction, and electromagnetic interaction) have been recently proposed. For example, band gap tunability can be achieved using magnetoactive [57] or photo-responsive [58] materials, wave guiding can be attained in piezoelectric phononic crystals [59], and complex rectifying devices can be realized in acoustics using piezoelectric membranes [60]. The use of programmable switches based of bonded piezoelectric patches has been exploited for dynamic reconfigurable topological waveguiding [61]. Mechanically triggered variations have also been proposed, e.g., the use of prestress to tune wave band gaps [62] [63,64], also realized experimentally. However, in this case, strains or deformations need to be relatively high to obtain significant changes, with correlated damage risks. One way to address this problem is to use soft materials (e.g., elastomers), with the additional complication of large deformations, nonlinearity, buckling and instabilities [65–67]. An alternative to this could be to use auxetic metamaterials. This enables large volumetric deformations at relatively low values of stress due to the small bulk modulus. An example of this has been recently presented to demonstrate band gap tunability for both elastic and acoustic metamaterials [68].

In this paper, we present the design and numerical analysis of an auxetic metamaterial, whose propagation properties can be reversibly tuned by applying external pre-strains. The paper is organized as follows. In Section 2, we discuss the dispersion properties of an auxetic medium with cuts arranged in a hexagonal pattern, for which the openings of band gaps in correspondence of broken Dirac cones are obtained by perturbing the cut lengths. The topological properties of the system are quantitatively demonstrated by calculating the Berry curvature and the valley Chern number. In addition, we perform numerical simulations to show the occurrence of robust wave propagation at the interface between two sub-domains consisting of cells with different perturbations of the cut lengths. In Section 3, we introduce the effect of a uniform pre-strain considering that the material is nonlinear. We show that the presence of an externally applied uniaxial strain can make the topologically protected interfacial wave disappear or be moved to a different frequency. An additional computation shows the robustness of the proposed effect where, as a consequence of the application of a non-uniform pre-strain, wave localization is lost in a finite region, but naturally reappears beyond it, where the pre-strain is absent. In Section 4, we present some concluding remarks, and in the Appendices, we provide additional details.

## 2. Auxetic topological metamaterial

### 2.1 Unit cell definition and dispersion properties

Based on previous work [22,69,70], we consider a thin slab with oriented cuts, consisting of hexagonal unit cells that are periodic in the directions defined by the lattice vectors $a_1 = (1, 0)^T$ and $a_2 = (1/2, \sqrt{3}/2)^T$ in the plane *x-y*, as shown in Fig. 1a. We start from a "symmetric" unit cell with cuts of equal length, which displays $C_6$ symmetry (Fig. 1b). The cuts are all rotated by the same relative angle $\theta = \pi/4$ with respect to the cell edges, as shown in Fig. 1b. To break the symmetry and open a topological band gap, we modify the hexagonal unit cells by shortening three cuts (II, IV, VI) out of six, to obtain a $C_3$ symmetry (Fig. 1c). The length of the shorter cuts in the "non-symmetric" unit cell is given by $a+b-2l'$ (see Fig. 1c for the symbols).

The auxetic behaviour of both geometries (symmetric and non-symmetric) is evaluated considering a periodic elementary cell and numerically determining the effective Poisson's ratio and Young's modulus by applying macroscopic strains and periodic boundary conditions [22] (the computation is performed in the linear elastic regime and the matrix material is assumed to be isotropic). The hexagonal $C_6$ and the trigonal $C_3$ symmetries ensure the in-plane isotropy of the effective behaviour [71]. The isotropy is also demonstrated by the isofrequency contours reported in Appendix A. For the symmetric cell, we obtain an effective Poisson's ratio which is very close to the lower limit for isotropic media of -1, i.e., $\nu_{eff} = -0.98$. For the non-symmetric cell, there is an increase in the effective Poisson's ratio, which, for the parameters considered in this work (see below), is given by $\nu_{eff} = -0.36$, implying that the material is still auxetic.

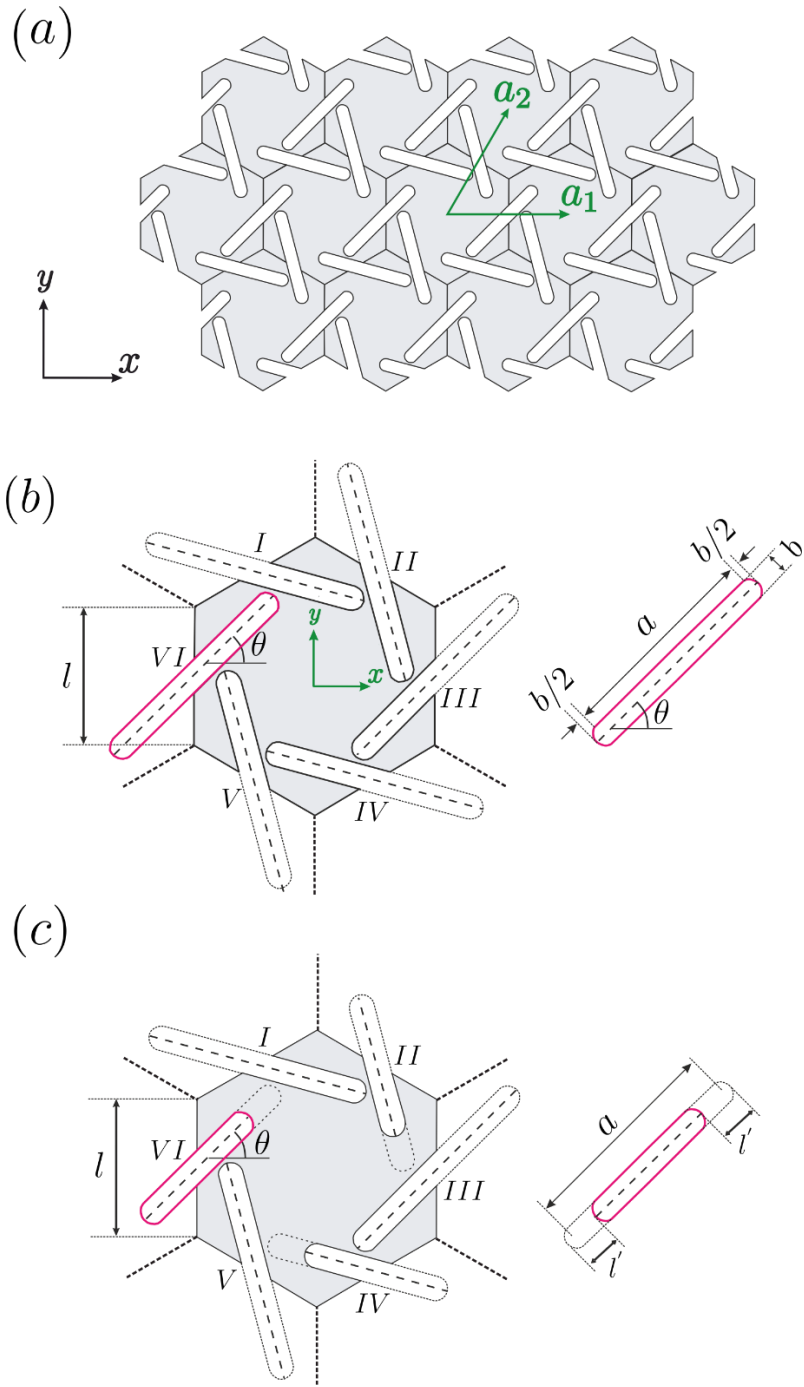

Figure 1: Schematic representation of the porous metamaterial in symmetric and non-symmetric configurations. (a) The hexagonal lattice composed of the periodic unit cells along lattice vectors *a₁* and *a₂*. (b) The geometrical details of the symmetric unit cell (all the cuts have the same length

*a+b). (c) The geometrical details of the non-symmetric unit cell after symmetry breaking (the cuts II, IV and VI have length a+b-2l').*

We begin by computing the dispersion properties and constructing band diagrams for the thin solid (3D) periodic systems characterized by the symmetric and non-symmetric unit cells presented in Fig. 1. These relations are obtained by enforcing periodic Bloch-Floquet boundary conditions on the edges of the unit cell and scanning the contour of the corresponding irreducible Brillouin zone, with high-symmetry points given by $\varGamma \equiv (0,0)$, $K \equiv (4\pi/3L, 0)$, and $M \equiv (\pi/L, \pi/L\sqrt{3})$, where $L = \sqrt{3}l$ is the distance between the centres of two adjacent unit cells. The geometrical parameters of the unit cell are taken as follows: cut length $a = 13$ mm, cut width $b = 1.6$ mm, hexagonal cell size $l = 9$ mm and cut length reduction $l' = 2.5$ mm. Various values of thickness are considered, ranging from $h = 0.5$ mm to $h = 2.0$ mm, with increasing steps of 0.5 mm. For material properties, we consider an elastomeric material with Lamé moduli $\mu = 283$ kPa and $\lambda = 806$ kPa and density $\rho = 970$ kg/m³ [72]. Finite element models are obtained by discretizing the unit cell using hexahedral solid elements with a characteristic length of 0.25 mm. For each case, a polarization metric in terms of displacement components, given by

$$p = \int_V |u_z|^2 \, dV \, / \int_V \left( |u_x|^2 + |u_y|^2 + |u_z|^2 \right) dV , \qquad (1)$$

is computed, which yields a value of 1 for purely out-of-plane wave modes and 0 for purely in-plane modes. Figure 2a shows that an increase in the thickness of the unit cell greatly influences the out-of-plane modes ($p \approx 1$), while the in-plane modes ($p \approx 0$) remain practically unchanged. Also, in-plane and out-of-plane modes are fully decoupled due to the symmetry of the unit cell along the out-of-plane axis. The band structure of the unit cell with equal cuts exhibits Dirac cones for the in-plane polarization in correspondence with the *K* point (Fig. 2a, top row). The Dirac point represents a single contact point between two dispersion surfaces, where two modes become degenerate [55]. The band diagrams indicate a first (second) Dirac cone at the frequencies of 159 Hz (338 Hz) and 162 Hz (340 Hz) for the thickness values of $h = 0.5$ and $h = 2.0$ mm, respectively. The band diagrams corresponding to $h = 1.0$ and $h = 1.5$ mm are omitted for the sake of brevity.

To obtain a topologically protected state in the system, it is necessary to break the $C_6$ symmetry, reducing it to $C_3$ symmetry, as detailed in Fig. 1c. For this non-symmetric unit cell, band structure calculations (Fig. 2a, bottom row) confirm that two topological band gaps appear, which are computed between the second and third (fifth and sixth) bands, at the frequency ranges of 275-336 Hz (517-535 Hz) and 277-338 Hz (521-541 Hz), respectively, for the thickness value of $h = 0.5$ ($h = 2.0$ mm). The opening of a total band gap in proximity of the broken Dirac cone has been observed only for an auxetic metamaterial. In Appendix B, we show that this effect does not occur in the corresponding non-auxetic system.

The negligible influence of the unit cell thickness on the in-plane modes in the dispersion diagrams suggests that a two-dimensional plane stress model is sufficient to capture the representative dynamics of the system associated with in-plane modes. The dispersion relations for the in-plane polarized modes are thus computed using quadrilateral elements for both symmetric and non-symmetric unit cells (Fig. 2(b), top and bottom rows, respectively), indicating a good correlation between the solid (filled circles) and plane stress (red hollow circles) models. Most notably, the

Dirac cones are formed at 156 Hz and 335 Hz for the symmetric unit cell, while the corresponding band gaps are opened between 273-335 Hz and 515-532 Hz for the non-symmetric unit cell .

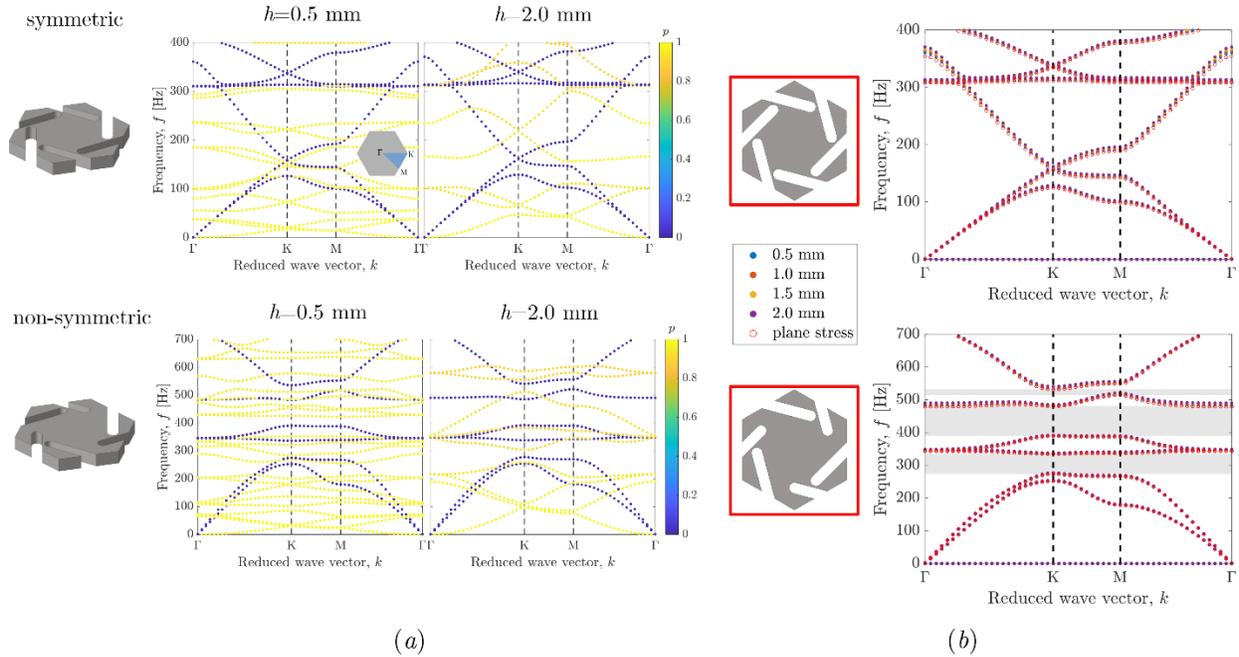

Figure 2: Band diagrams computed for the (a) solid models of the unit cells with increasing values of thickness h for the symmetric (top row) and non-symmetric (bottom row) unit cells. Colour scales refer to polarization values (0 for in-plane, 1 for out-of-plane wave modes). The negligible influence of thickness on in-plane modes suggests the use of a plane stress model as a reasonable approximation, which is confirmed by (b) the comparison between the band diagram of solid models (filled circles) and a plane stress model (red hollow circles). Total band gaps are marked in grey.

2.2    Determination of the valley Chern number for the non-symmetric unit cell

We now study the topological properties of the perturbed structure (non-symmetric unit cell), characterized by two types of cuts with different lengths. To ensure that the considered perturbed structure can support topologically protected edge modes, we calculate the valley Chern number and the map of the Berry curvature in the reciprocal space.

We focus the attention on the second and third dispersion surfaces, between which a band gap appears when the symmetry is broken (see Fig. 2b). For each dispersion surface, we compute the eigenmode $U(k) = (U_x(k), U_y(k))^T$ for different values of the wave vector $k = (k_x, k_y)^T$ in the reciprocal space. We notice that the frequency of the eigenmode at any value of $k$ also depends on the choice of the dispersion surface.

The *valley Chern number* is defined as [49,73]

$$C_v = \frac{1}{2\pi} \int_A \Omega(k) \mathrm{d}^2 k, \tag{2}$$

where $\Omega(k)$ represents the *Berry curvature* and $A$ denotes a small area around a "valley". In fact, since the considered perturbed system does not break time-reversal symmetry, the integration of the Berry curvature over the whole Brillouin zone is null; nonetheless, the Berry curvature is localized at specific positions ("valleys") of the reciprocal space, in particular around $K$ and $K'$ points (see Fig. 3), so that the integration of $\Omega(k)$ over a small area in correspondence of these points is different from zero.

For each dispersion surface, we calculate the Berry curvature $\Omega(k)$ over a prescribed domain of the reciprocal space using the following procedure. First, we subdivide the domain into small rectangular patches, whose vertices are denoted by $P_1$, $P_2$, $P_3$ and $P_4$ (taken in the counter-clockwise direction). For each patch, we extrapolate the eigenvectors at the $k$-values corresponding to the four vertices. Then, we compute the Berry curvature as follows [74]:

$$\Omega(\boldsymbol{k}) = -\mathrm{Im}\left[\log\left(\frac{\langle \boldsymbol{U}(P_1)|\boldsymbol{U}(P_2)\rangle \langle \boldsymbol{U}(P_2)|\boldsymbol{U}(P_3)\rangle \langle \boldsymbol{U}(P_3)|\boldsymbol{U}(P_4)\rangle \langle \boldsymbol{U}(P_4)|\boldsymbol{U}(P_1)\rangle}{\langle \boldsymbol{U}(P_1)|\boldsymbol{U}(P_1)\rangle \langle \boldsymbol{U}(P_2)|\boldsymbol{U}(P_2)\rangle \langle \boldsymbol{U}(P_3)|\boldsymbol{U}(P_3)\rangle \langle \boldsymbol{U}(P_4)|\boldsymbol{U}(P_4)\rangle}\right)\right], \quad (3)$$

where

$$\langle \boldsymbol{U}(P_i)|\boldsymbol{U}(P_j)\rangle = \frac{1}{2\pi}\int_S \boldsymbol{U}^*(\boldsymbol{k}(P_i)) \cdot \boldsymbol{U}(\boldsymbol{k}(P_j))\mathrm{d}S. \quad (4)$$

In the formula above, the symbol * indicates the complex conjugate and $S$ represents the area of the periodic cell.

The colour maps of the Berry curvature associated with the second and third dispersion surfaces are shown in Fig. 3a and 3b, respectively. Each point represents the centroid of a patch in the shape of a parallelogram of sizes $0.005 \frac{4}{\sqrt{3}} \frac{\pi}{L}$ in the $k_x$- and $k_y$-direction. As expected, we note that the Berry curvature exhibits peaks in correspondence of the characteristic points of the reciprocal space, namely $K$ /$K'$.

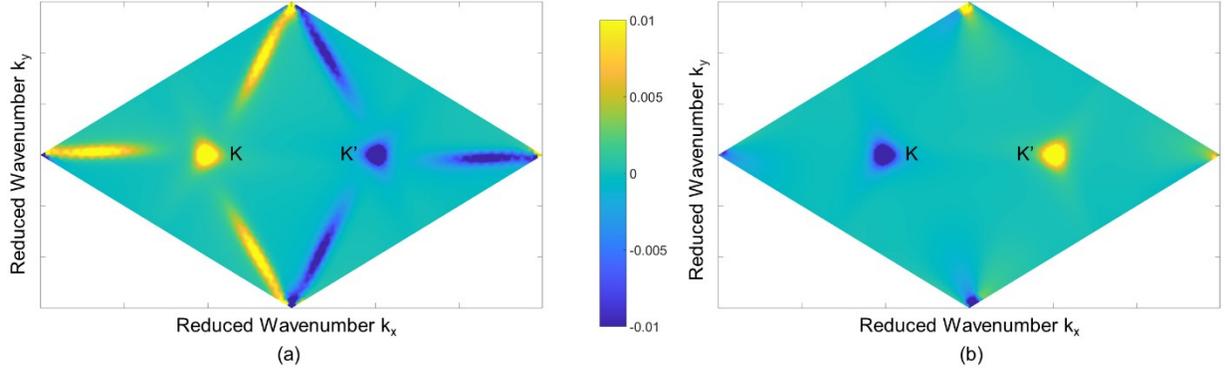

*Figure 3: Values of the Berry curvature Ω in the reciprocal space for the (a) second and (b) third dispersion surface of the perturbed structure (non-symmetric unit cell, Fig. 1c). The characteristic points K and K' are also reported.*

Using Eq. (2), we can evaluate the valley Chern number at the above-mentioned characteristic points. For both dispersion surfaces, $C_v = 0$ at $\Gamma$ (vertices of the parallelograms). On the other hand, for the second dispersion surface, $C_v$ = -0.49525 at $K$ while $C_v$ = +0.49581 at $K'$. Opposite values are found for the third dispersion surface. This confirms that the considered system with broken symmetry allows the generation of topologically protected edge modes [73].

### 2.3   Topological edge modes

The existence of valley edge modes can be verified by considering a supercell consisting of two distinct domains, constituted by the non-symmetric unit cells labelled as A (top half) and B (bottom half), comprising a total of 16-unit cells separated by an interface in the middle (Fig. 4a). The unit cell B is obtained by rotating the unit cell A by 60° in the counterclockwise direction. The resulting supercell can be analysed as a structure which is periodic in the *x*-direction and has a finite length in the *y*-direction, with free boundary conditions at the top and bottom edges of the supercell. The corresponding dispersion diagram can be computed by scanning the first Brillouin zone, delimited, in this case, by the high-symmetry points $\Gamma(0,0)$ and $X(\pi/L, 0)$.

Fig. 4 shows the resulting band diagram in three distinct frequency ranges for a clearer visualization, namely 260-340 Hz (Fig. 4b), 335-355 Hz (Fig. 4c), and 380-480 Hz (Fig. 4d). In each case, the bands which represent localized interface modes are tracked through the modal assurance criterion and highlighted in red. The corresponding wave modes, labelled as $I_1$ through $I_5$, are also computed at the *X* point of the reciprocal lattice and shown with the colour bar representing displacement amplitude values. The frequencies of the represented wave modes at the *X* point are 325 Hz, 345 Hz, 347 Hz, 418 Hz, and 453 Hz, respectively.

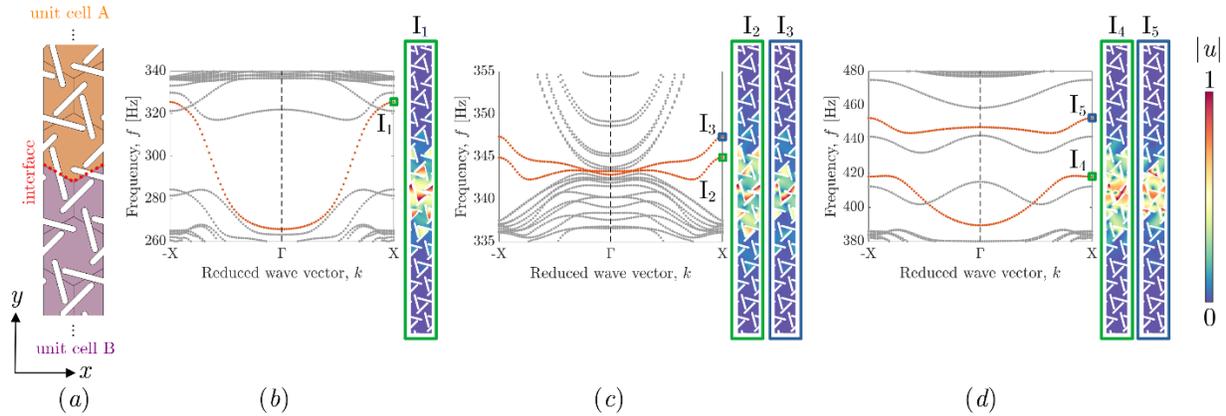

*Figure 4: Band diagrams of the supercell. (a) The supercell is obtained as a finite strip composed of 16 unit cells, with the bottom and top halves constituted by non-symmetric unit cells A and B, respectively, rotated by 60° with respect to each other. The resulting band diagrams are shown for distinct frequency ranges, i.e., (b) 260-340 Hz, (c) 335-355 Hz, and (d) 380-480 Hz, with red bands representing topologically localized modes $I_1$ through $I_5$, which show most significant relative displacements at the interface of the two sub-domains of the supercell.*

## 2.4 Propagation of topologically protected modes

The topological waveguiding properties can be demonstrated by constructing a finite-sized sample with an interface between the unit cell regions A and B determining the preferential direction of ideally lossless energy propagation. For this purpose, a finite structure composed of 22 x 34 unit cells is designed, with a sharp corner of 60º between distinct segments of the interface. This structure is presented in Fig. 5a, where the interface is highlighted in red and the point where an in-plane excitation is applied in the horizontal direction is also shown. Frequency-domain steady-state analyses are performed using signals with single frequencies corresponding to the excitation of isolated interface modes. The wave mode corresponding to $I_1$, for instance, ranges from approximately 265 Hz (in Γ) to 325 Hz (in X) (Fig. 4b) For this reason, one of the excitation frequencies is chosen as 325 Hz. The second excitation frequency is taken as 415 Hz, which corresponds to mode $I_4$ (Fig. 4d). The results in Fig. 5b and Fig. 5c demonstrate the concentration of energy at the interface between unit cells. Additionally, a transient analysis is performed considering a sine "burst" signal obtained by modulating 50 sinusoidal cycles with a Hanning window (inset of Fig. 5a) to centre the frequency content of the applied input and avoid energy leakage. The resulting absolute displacements are shown for distinct time instants in Fig. 5d and Fig. 5e, respectively, at 95 and 154 ms. The propagation of energy is thus shown to be concentrated at the interface between the two unit cell types, with a steep decay moving away from the interface. The presence of the sharp corner does not preclude the propagation of the localized mode and no scattering effects are observed. The propagation in the transient regime is also reported in the Supplementary Video.

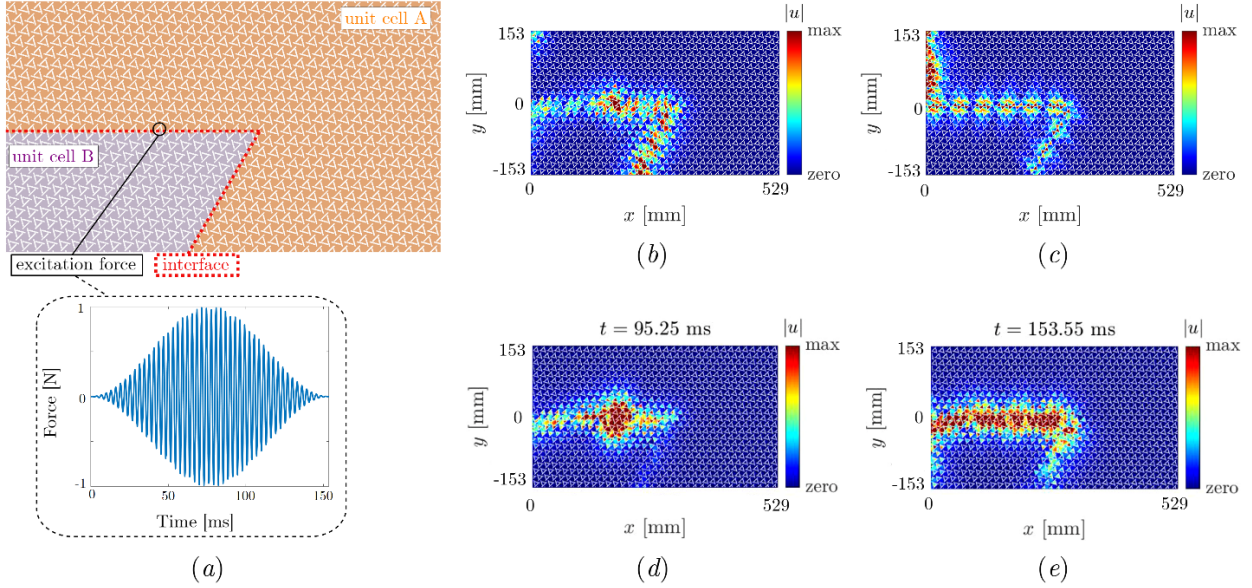

*Figure 5: Topological waveguiding functionality. A finite structure comprised of 22 x 34 unit cells, with the interface highlighted in red, is excited in the horizontal direction at the centre of the region indicated by the black circle. The concentration of energy at the interface region is verified by frequency-domain steady-state analyses with input frequencies of (b) 325 Hz and (c) 415 Hz. Transient analysis with applied sine bursts centred at 325 Hz (inset of part (a)) for time instants (d) 95 ms and (e) 154 ms, shows excitation of a localized mode, which propagates along the interface.*

## 3. Tunability of topologically protected wave propagation

### 3.1 Band gap manipulation applying external strain

The properties of the discussed auxetic metamaterial can become tunable by applying quasi-statically pre-strains leading to a nonlinear elastic response. To illustrate this, we choose a Neo-Hookean hyperelastic model [75,76], which describes the reversible nonlinear behaviour of materials like polymers or rubbers for large deformations. The hyperelastic strain energy function has the form

$$W(\boldsymbol{F}) = \frac{\mu}{2}(J_1 - 3) - \mu \ln(J_{el}) + \frac{\lambda}{2} [\ln(J_{el})]^2, \qquad (5)$$

where $\boldsymbol{F} = \nabla \boldsymbol{x}$ is the deformation gradient, with $J_1 = \mathrm{tr}(\boldsymbol{F}^T \boldsymbol{F})$ and $J_{el} = \det(\boldsymbol{F})$. In the following we will apply a uniaxial stretch $F_{yy}=1+\varepsilon_{yy}$ of moderately large amplitude and present results as functions of the strain component $\varepsilon_{yy}$.

Simulations are performed using the structural mechanics and nonlinear elasticity modules of *COMSOL Multiphysics*. The finite element analysis is carried out by means of a coupled "stationary" and "eigenfrequency" study, whereby first a large quasistatic pre-strain (with values up to 0.2) is applied to the (non-symmetric) unit cell including material and geometrical nonlinearity, and subsequently the corresponding band structures of the deformed unit cells are determined. Results are presented in Fig. 6b and Fig. 6c, where the deformed configurations of the unit cells are depicted, with the local von Mises Cauchy stress distribution represented in colour scale, for $\varepsilon_{yy} = 0.1$ and $\varepsilon_{yy} = 0.2$, respectively. Cauchy stresses reach values of approximately 0.2 MPa and 0.4 MPa, respectively, concentrating at slender material portions at the tip of the cuts, where deformations are greatest.

The band structures of the deformed unit cells of the nonlinear elastic material undergo significant changes compared to those for the unloaded ones. Fig. 6b and Fig. 6c show that in the presence of a pre-strain of 0.1 and 0.2, respectively, the first topological band gap closes, thus eliminating the possibility of generating topologically protected modes at this frequency, while the second band gap is preserved but shifted to higher frequencies (from 382 – 452 Hz for $\varepsilon_{yy} = 0$ to 428 – 488 Hz for $\varepsilon_{yy} = 0.2$). This introduces the possibility of different manipulation of waveguiding effects at the two considered frequencies. In the case of the lower frequency range, waveguiding can be eliminated by application of a sufficiently large uniaxial external strain, while in the higher range it can be tuned to higher frequencies. It should be noticed that the effect is mainly due to geometrical nonlinearities, since at the chosen strain levels material nonlinearity has a small influence on results. This has been verified numerically by "turning off" material nonlinearity while maintaining geometrical nonlinearity (not shown here for brevity). Results for the band diagrams in this case are very similar to those reported in Fig. 6b and Fig. 6c.

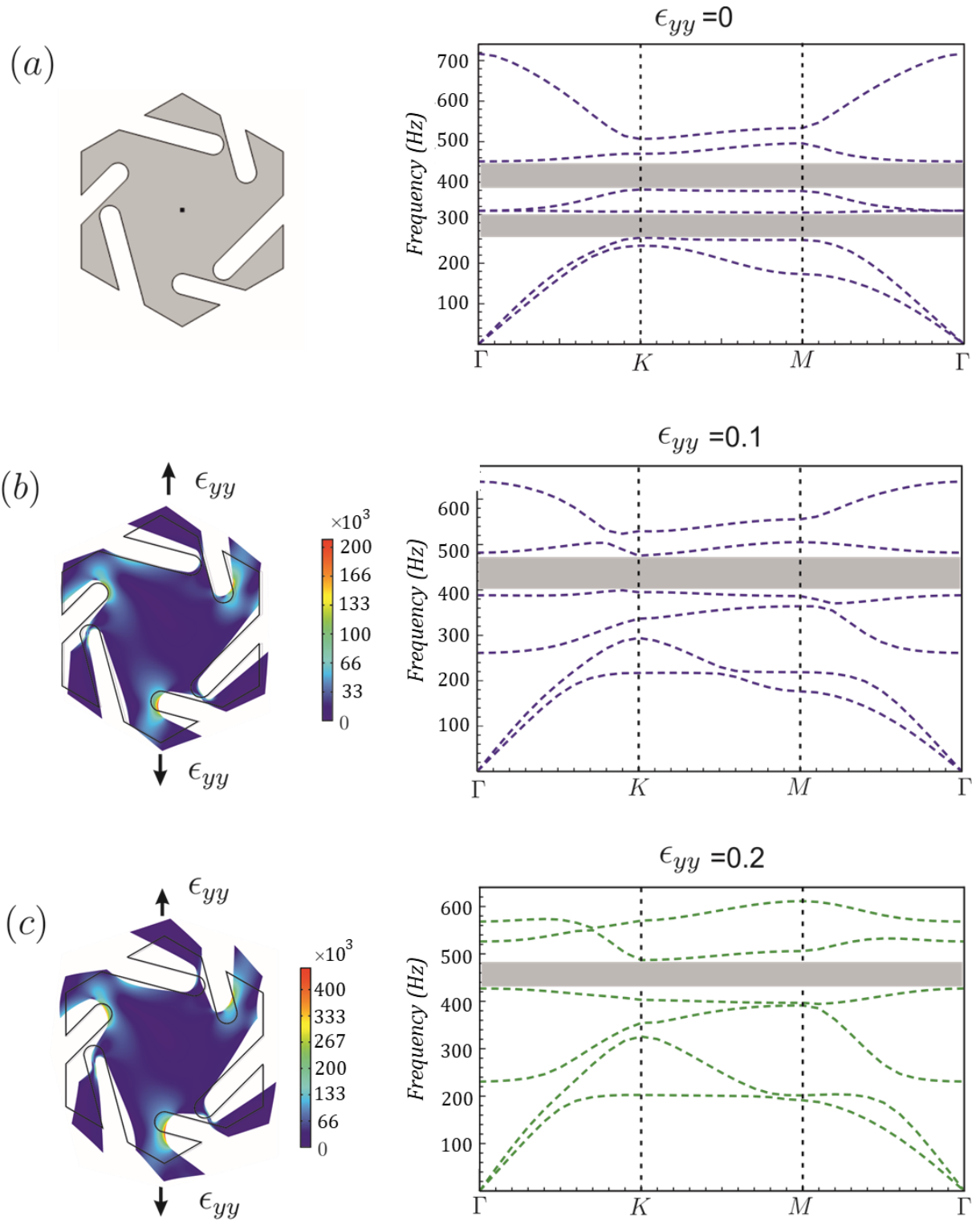

Figure 6: a) Undeformed configuration of the unit cell and corresponding band diagram; b) deformed configuration for $\varepsilon_{yy}=0.1$, with von Mises-Cauchy stresses represented in colour scale

*(in Pa), and corresponding band diagram; c) deformed configuration for $\varepsilon_{yy}=0.2$ and corresponding band diagram.*

For a deeper analysis of the phenomenon, a diagram illustrating how the lower two band gaps vary with the externally applied pre-strain is calculated. For this purpose, in Fig. 7 we show how the ranges of the two band gaps (coloured in grey) change with the applied pre-strain $\varepsilon_{yy}$. From the figure, it can be seen that the first band gap is closed at a strain that is slightly larger than 0.07. Conversely, the second band gap is maintained but shifted to higher frequencies as the value of pre-strain is increased; the frequency interval of the second band gap depends on the value of $\varepsilon_{yy}$.

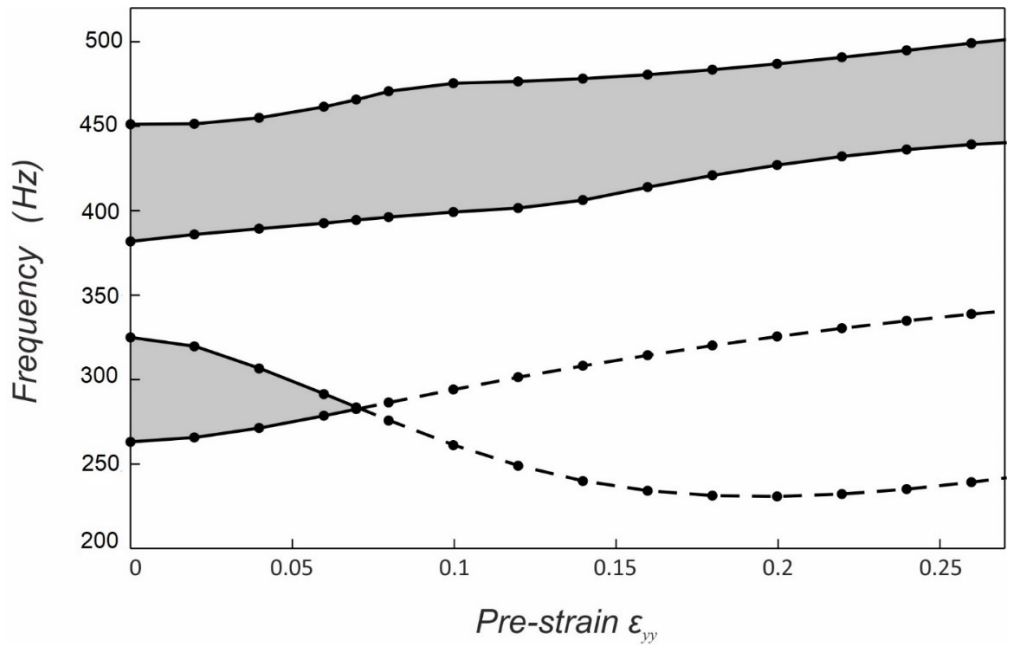

*Figure 7: Frequency ranges of the lower two band gaps (in grey colour) versus the imposed pre-strain $\varepsilon_{yy}$.*

### 3.2 Tuning waveguiding with external pre-strain

To verify the results emerging from band structure calculations and illustrate how these can be exploited in applications requiring control of waveguiding, we present a configuration which includes 50 unit cells in the $x$ direction and 32 unit cells in the $y$ direction (Fig. 8a). We perform frequency domain simulations, exciting signals at the left side of the sample at a fixed frequency of 300 Hz, falling inside the lower topological band gap in the absence of prestrain (Fig. 8b) and corresponding to mode $I_1$ in Fig. 4. We evaluate the propagating wave field for increasing vertically applied quasi-static strain $\varepsilon_{yy}$ imposed on the entire upper edge of the specimen (the same strain is imposed on the lower edge of the domain). For small strain values (e.g., for $\varepsilon_{yy} <$ 0.05), displacements are well localized at the interface between the two domains, as observed in Fig. 4. With increasing vertical strain, there is a gradual weakening of the localization effect, and a transition to non-localized propagation occurs at about $\varepsilon_{yy} = 0.075$. This corresponds to a strain value for which the band gap centred at 300 Hz is closed (Fig. 8b), which thus results in the possibility of mode conversion effects and partial transfer of energy to other modes. As discussed, the variation in the topological band gaps inducing the disappearance of waveguiding effects is mainly due to the presence of geometrical nonlinearity in the deformation of the auxetic structure.

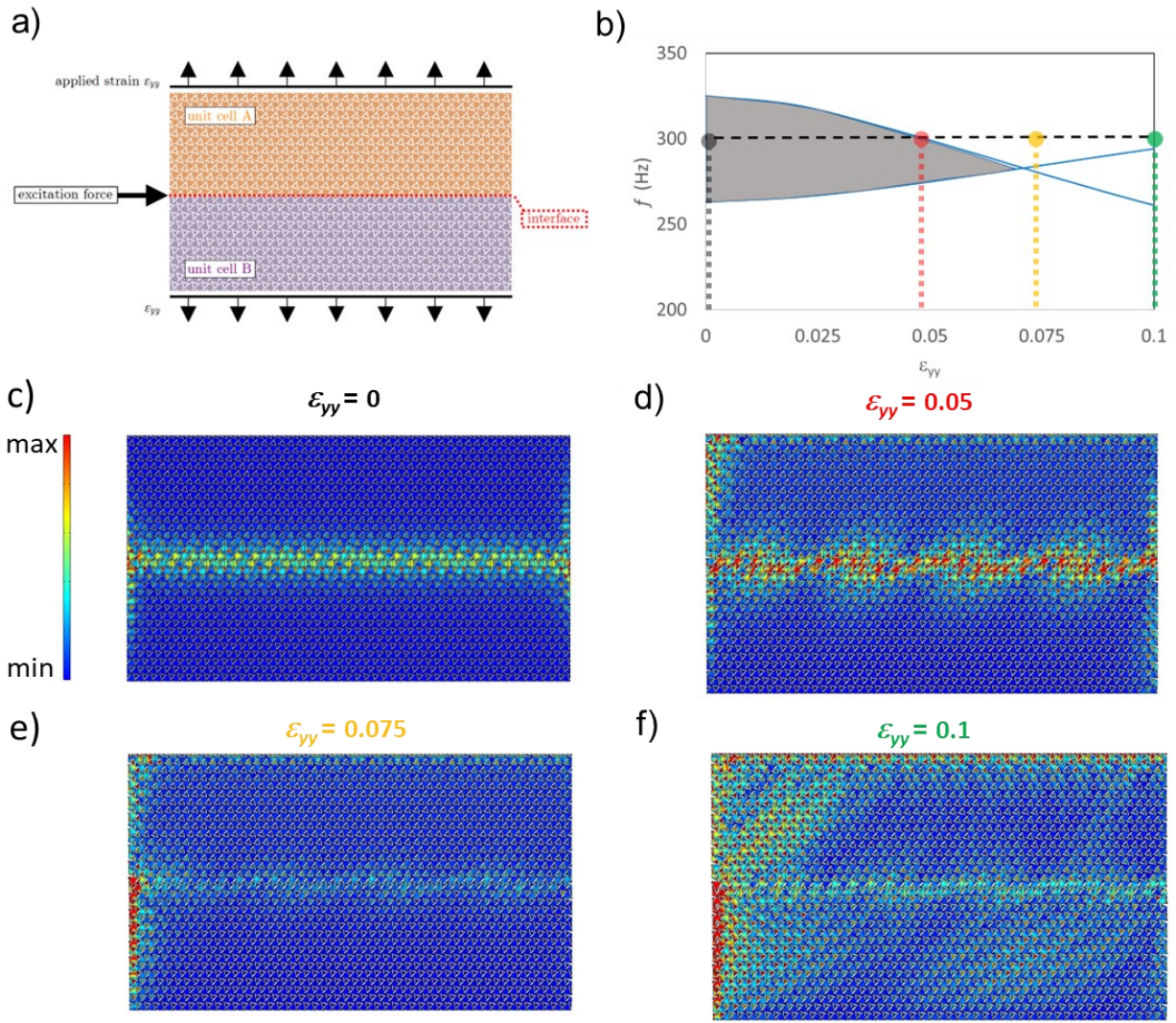

*Figure 8: Tuning of waveguiding by applying an external pre-strain $\varepsilon_{yy}$. (a) Schematic of the FEM model for frequency domain simulations. b) Detail of the plot in Fig. 7, focusing on the lower band gap, highlighting the frequencies and pre-strains considered in the simulations. Displacement fields for f=300 Hz and c) $\varepsilon_{yy}$ =0, d) $\varepsilon_{yy}$ =0.05, e) $\varepsilon_{yy}$ =0.075, and f) $\varepsilon_{yy}$ =0.1, illustrating progressive loss of wave localization for increasing pre-strain.*

On the other hand, the application of an external vertical strain can also be exploited to generate a frequency shift in the waveguiding effects of the localized modes at the interface between the two domains. This takes place for wave propagation at the higher of the two considered frequency ranges in the unit cell band diagrams, illustrated in Fig. 9a. In this case, application of an external vertical pre-strain induces a shift of waveguiding effects to higher frequencies: $f = 440$ Hz for $\varepsilon_{yy}=$ 0.05 (Fig. 9b), 450 Hz for $\varepsilon_{yy}= 0.1$ (Fig. 9b), and 462 Hz for $\varepsilon_{yy}= 0.15$ (Fig. 9c), respectively. Notice that in this case, the upper and lower band gap bounds overestimate the range in which protected waveguiding occurs, since mode $I_4$ only covers a relatively small frequency range within the band gap (see Fig. 4).

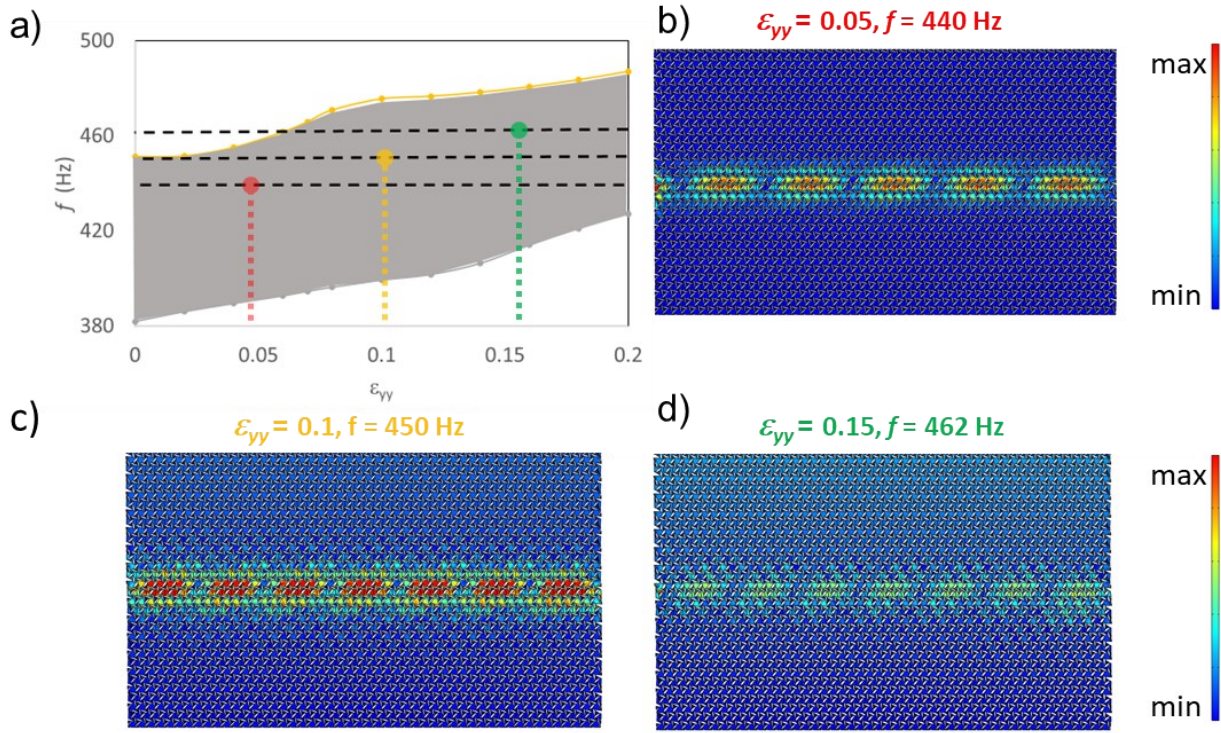

Figure 9: a) Detail of the plot in Fig. 7 highlighting the frequencies and pre-strains considered in the simulations. Displacement fields showing wave localization at increasing frequencies in the

*presence of an increasing vertical pre-strain $\varepsilon_{yy}$: b) f = 440 Hz for $\varepsilon_{yy}$ =0.05, c) f =450 Hz for $\varepsilon_{yy}$ =0.1, d) f =462 Hz for $\varepsilon_{yy}$ =0.15.*

The effect of a quasistatic pre-strain field on the localization of wave propagation can also be exploited to modify the waveguiding properties of a structure locally, in a confined region. This is schematically illustrated in Fig. 10 on a specimen composed of 75 unit cells in the *x* direction and 32 unit cells in the *y* direction, in which an external vertical pre-strain is applied on a limited length in the centre of the sample (1/3 of the length), as highlighted in Fig. 10a. In the specific example, waves are excited at 300 Hz, corresponding to the localized waveguiding of mode $I_1$, and a pre-strain value of $\varepsilon_{yy}$ = 0.1 is chosen. As shown in Fig. 10b, the applied pre-strain disturbs the localization of the wave propagation only along a restricted length, of the order of the distance along which the pre-strain is applied, thus allowing wave control in targeted spatial regions of the sample. Very similar results are obtained in other frequency regions, targeting other localized modes discussed previously.

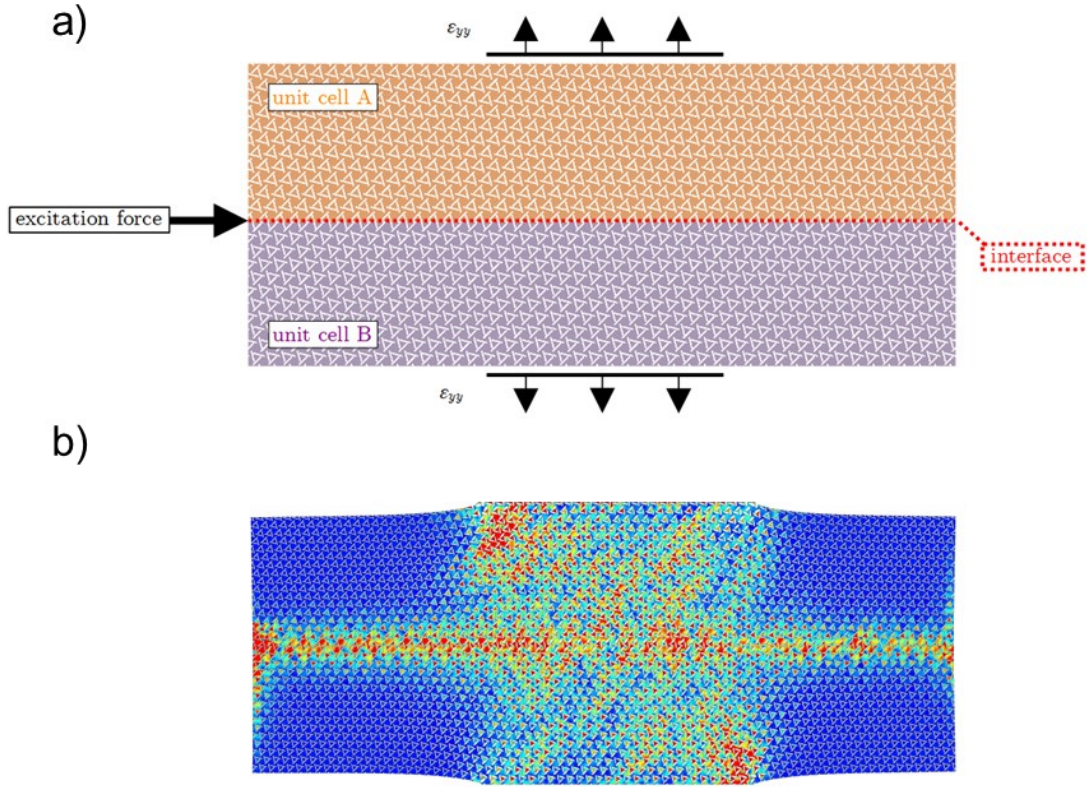

*Figure 10: (a) Schematic of the selected model and loading scenario. (b) Localized wave-guiding for $\varepsilon_{yy} = 0.1$ at 300 Hz. Localized mode waveguiding is inhibited only in correspondence with the spatial location of the application of pre-strain.*

## 4. Conclusions

In conclusion, we have presented a numerical study to demonstrate the possibility of creating dynamically reversibly tunable topological metamaterials applying pre-strains, exploiting their auxeticity and nonlinearity.

The proposed unit cell design is quite simple, and only requires the introduction of oriented cuts arranged in a hexagonal pattern within a thin sheet of material. The structure has interesting properties, both quasistatic and dynamic, in that it exhibits auxetic behaviour with an extremely negative Poisson's ratio; at the same time, the band structure for both in-plane and out-of-plane modes admits various Dirac cones, which can be exploited for the creation of topological valley modes. We have shown that the in-plane modes are relatively independent of thickness effects, allowing a 2D treatment of the problem without loss of generality.

The breaking of symmetry of the unit cells to remove degeneracy is obtained by changing the length of alternate cuts, also allowing parametric design of the desired dispersion properties. We have shown that this feature only occurs in the presence of cell auxeticity. After symmetry breaking and construction of macrocells and/or large lattices with interface regions, we have shown good wave localization at the interface and lossless (scatter-free) propagation at sharp edges.

The novelty compared to previous designs and conceptual experiments is that, in this case, the lossless waveguiding properties can be manipulated by the simple application of unidirectional quasi-static pre-strains of moderate amplitudes (typically below 10%), in which geometrical nonlinearity can be activated. This allows a controlled variation of the unit cell dispersion properties, with the removal or shifting of band gaps to higher frequencies, thus enabling the transition from localized waveguiding to nonlocalized propagation. This modification can also be generated in limited portions of the propagation domain, by applying strain that is variable in space or time. Auxeticity plays a role in this process in two ways: first, it allows to open band gaps in the specific structure under study, which is the starting point for topologically protected waveguiding; second, it allows to achieve particular strain states in the material subjected to uniaxial loading, e.g. a biaxial expansion which helps to conserve the unit cell proportions and thus shift modes and band gaps in frequency. A future detailed study of how strain states correlate to band gap modification and topological protection in this type of system is therefore potentially of great interest.

Given the relative simplicity of the design, these auxetic tunable topological metamaterials should be amenable to fabrication in soft polymeric or elastomeric samples with standard techniques, and the required material/geometrical nonlinearity could be generated at relatively small strains, allowing for an experimental realization of proof-of-concept experiments, and further extending the possibilities in the fast-growing field of tunable metamaterials.

**Acknowledgments**

The authors are grateful to Prof. Massimiliano Fraldi for useful discussions on Appendix A. M. Mor., V.F.D.P., M. Min., N.M.P., A.S.G. and F.B. are supported by the European Commission H2020 FET Open "Boheme" grant no. 863179. G.C. and M.B.'s work has been performed under the auspices of GNFM-INDAM.

**Appendix A – Isofrequency contours**

Here, we report the isofrequency contours for the symmetric structure shown in Fig. 2a. The contours have been computed at the low frequencies $f$ = 10, 20 Hz and are reported in Fig. A1a and A1b for the first and second dispersion surfaces, respectively. The "circular" contours indicate the isotropy of the effective behaviour. The same result was obtained in [22] for $C_6$ symmetric microstructure, but stemming from the too restricting assumption of orthotropy for the considered model.

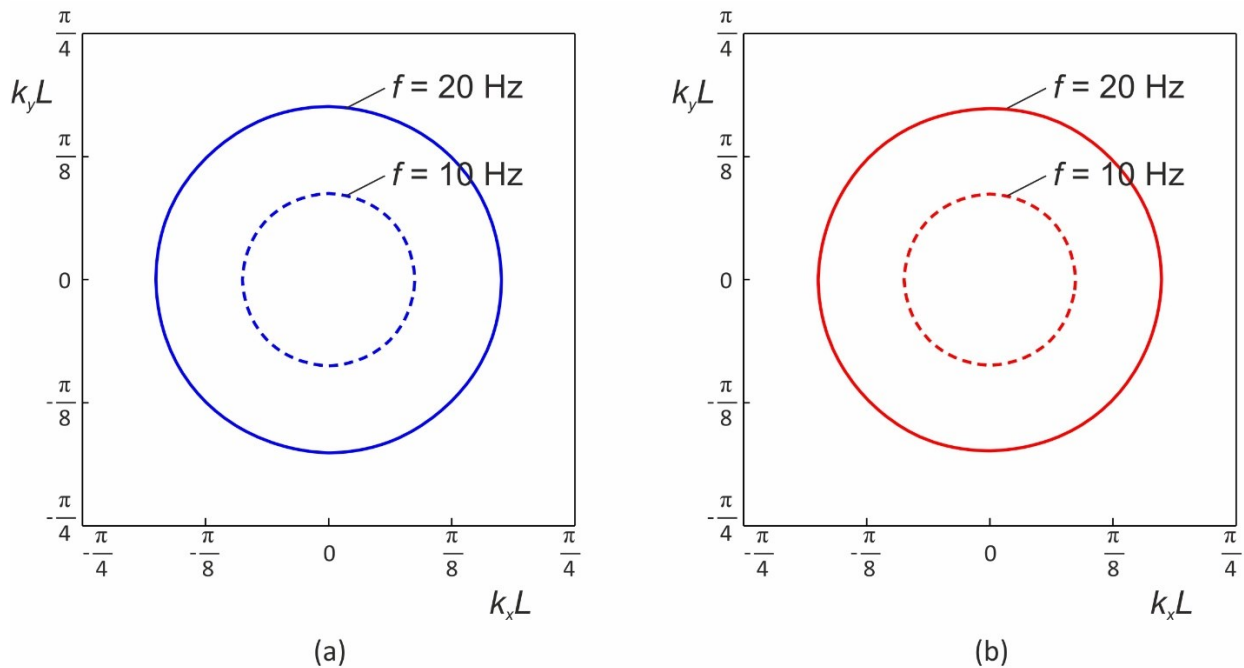

*Figure A1: Isofrequency contours for the (a) first and (b) second dispersion surface of the symmetric unit cell, computed at two different values of the frequency (i.e., $f$ = 10, 20 Hz).*

## Appendix B – Influence of auxeticity on dispersion

Here, we present the results of a parametric analysis, where the effective Poisson's ratio of the symmetric unit cell in Fig. 1b is varied by changing the ratio $a/l$ between the cut length and the hexagonal cell size, keeping the angle $\theta$ fixed and equal to $\theta = \pi/4$.

The dependence of the macroscopic Poisson's ratio $\nu_{eff}$ on the ratio $a/l$ was determined for different values of $\theta$ in [69]. The outcomes for the case $\theta = \pi/4$ are reported in Fig. B1a. It can be seen that $\nu_{eff}$ monotonically decreases as the ratio $a/l$ is increased, varying from positive to negative values. In particular, for $a/l = 0.6$, $\nu_{eff} > 0$ (non-auxetic), while for $a/l = 1.0$ $\nu_{eff} < 0$ (auxetic).

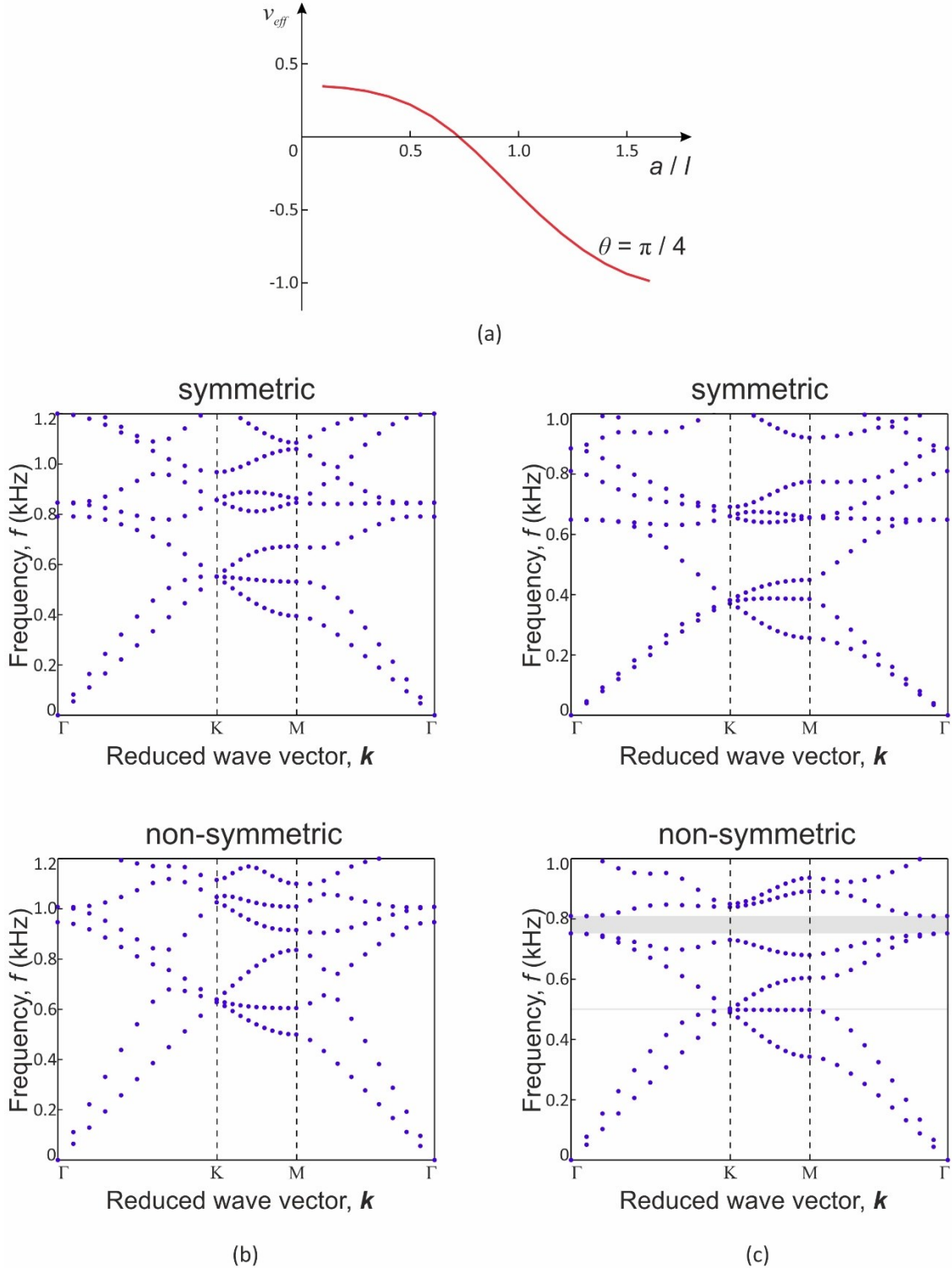

*Figure B1: (a) Effective Poisson's ratio $\nu_{eff}$ of the symmetric unit cell in Fig. 1b as a function of the ratio a/l for $\theta = \pi/4$ (data extrapolated from [69]). Dispersion diagrams for the symmetric*

*(top insets) and non-symmetric (bottom insets) unit cell for (b) a/l = 0.6 ($\nu_{eff} > 0$) and (c) a/l = 1.0 ($\nu_{eff} < 0$).*

In Fig. B1b we present the dispersion diagrams for the case *a/l* = 0.6, which corresponds to a positive value of the effective Poisson's ratio, for the symmetric unit cell (top figure) and for the non-symmetric unit cell (bottom figure) where the cut length reduction is $l' = 2.5$ mm as in Section 2.1. From the figure, it is apparent that the perturbation of the cut lengths does not lead to the opening of a band gap in the proximity of either the lower or upper broken Dirac cone. Consequently, topological wave propagation cannot take place in this situation, where $\nu_{eff} > 0$.

In Fig. B1c we show the band diagrams when the ratio *a/l* = 1.0, for which $\nu_{eff} < 0$. In this scenario, when the cut lengths are varied, a thin band gap appears in the neighborhood of the lower Dirac cone and another one of consistent width is opened at the upper Dirac cone. Accordingly, topological states can be obtained in this auxetic case.

From the above and further simulations (whose results are not reported here for brevity), we have observed that the possibility of creating topologically protected waves in the considered medium is closely related to its auxeticity.

# References

[1] R. H. Baughman, *Avoiding the Shrink*, Nature.

[2] R. Lakes, *Foam Structures with a Negative Poisson's Ratio*, Science **235** (4792), 1038-1040 (1987).

[3] K. K. Saxena, R. Das, and E. P. Calius, *Three Decades of Auxetics Research – Materials with Negative Poisson's Ratio: A Review*, Advanced Engineering Materials **18** (11), 1847-1870 (2016).

[4] L. Cabras and M. Brun, *A Class of Auxetic Three-Dimensional Lattices* Journal of the Mechanics and Physics of Solids **91**, 56-72 (2016).

[5] P. Mardling, A. Alderson, N. Jordan-Mahy, and C. L. Le Maitre, *The Use of Auxetic Materials in Tissue Engineering*, Biomaterials Science **8**, 2074-2083 (2020).

[6] H. M. A. Kolken, K. Lietaert, T. van der Sloten, B. Pouran, A. Meynen, G. Van Loock, H. Weinans, L. Scheys, and A. A. Zadpoor, *Mechanical Performance of Auxetic Meta-Biomaterials*, Journal of the Mechanical Behavior of Biomedical Materials **104**, 103658 (2020).

[7] J. I. Lipton, R. MacCurdy, Z. Manchester, L. Chin, D. Cellucci, and D. Rus, *Handedness in Shearing Auxetics Creates Rigid and Compliant Structures*, Science **360** (6389), 632-635 (2018).

[8] K. Chen, S. Fang, Q. Gao, D. Zou, J. Cao, and W. H. Liao, *Enhancing Power Output of Piezoelectric Energy Harvesting by Gradient Auxetic Structures*, Applied Physics Letters **120**, 103901 (2022).

[9] K. Chen, Q. Gao, S. Fang, D. Zou, Z. Yang, and W. H. Liao, *An Auxetic Nonlinear Piezoelectric Energy Harvester for Enhancing Efficiency and Bandwidth*, Applied Energy **298**, 117274 (2021).

[10] G. J. Chaplain, J. M. De Ponti, G. Aguzzi, A. Colombi, and R. V. Craster, *Topological Rainbow Trapping for Elastic Energy Harvesting in Graded Su-Schrieffer-Heeger Systems*, Physical Review Applied **14**, 054035 (2020).

[11] Q. Li, Y. Kuang, and M. Zhu, *Auxetic Piezoelectric Energy Harvesters for Increased Electric Power Output*, AIP Advances **7**, 015104 (2017).

[12] O. Duncan, T. Shepherd, C. Moroney, L. Foster, P. D. Venkatraman, K. Winwood, T. Allen, and A. Alderson, *Review of Auxetic Materials for Sports Applications: Expanding Options in Comfort and Protection*, Applied Sciences **8** (6), 941 (2018).

[13] S. C. Han, D. S. Kang, and K. Kang, *Two Nature-Mimicking Auxetic Materials with Potential for High Energy Absorption*, Materials Today **26**, 30-39 (2019).

[14] X. Ren, R. Das, P. Tran, T. D. Ngo, and Y. M. Xie, *Auxetic Metamaterials and Structures: A Review*, Smart Materials and Structures **27**, 023001 (2018).

[15] M. Kadic, T. Bückmann, R. Schittny, and M. Wegener, *Metamaterials beyond Electromagnetism*, Reports on Progress in Physics **76**, 126501 (2013).

[16]   T. Li, F. Liu, and L. Wang, *Enhancing Indentation and Impact Resistance in Auxetic Composite Materials* Composites Part B: Engineering **198**, 108229 (2020).

[17]   X. chun Zhang, C. chao An, Z. feng Shen, H. xiang Wu, W. gang Yang, and J. pan Bai, *Dynamic Crushing Responses of Bio-Inspired Re-Entrant Auxetic Honeycombs under in-Plane Impact Loading*, Materials Today Communications **23**, 100918 (2020).

[18]   L. Francesconi, A. Baldi, X. Liang, F. Aymerich, and M. Taylor, *Variable Poisson's Ratio Materials for Globally Stable Static and Dynamic Compression Resistance*, Extreme Mechanics Letters **26**, 1-7 (2019).

[19]   L. Francesconi, A. Baldi, G. Dominguez, and M. Taylor, *An Investigation of the Enhanced Fatigue Performance of Low-Porosity Auxetic Metamaterials*, Experimental Mechanics **60**, 93–107 (2020).

[20]   H. M. A. Kolken, A. F. Garcia, A. Du Plessis, C. Rans, M. J. Mirzaali, and A. A. Zadpoor, *Fatigue Performance of Auxetic Meta-Biomaterials*, Acta Biomaterialia **126**, 511-523 (2021).

[21]   R. Gatt, L. Mizzi, J. I. Azzopardi, K. M. Azzopardi, D. Attard, A. Casha, J. Briffa, and J. N. Grima, *Hierarchical Auxetic Mechanical Metamaterials*, Scientific Reports **5**, 8395 (2015).

[22]   M. Morvaridi, G. Carta, F. Bosia, A. S. Gliozzi, N. M. Pugno, D. Misseroni, and M. Brun, *Hierarchical Auxetic and Isotropic Porous Medium with Extremely Negative Poisson's Ratio*, Extreme Mechanics Letters **48**, 101405 (2021).

[23]   E. Baravelli and M. Ruzzene, *Internally Resonating Lattices for Bandgap Generation and Low-Frequency Vibration Control*, Journal of Sound and Vibration **332**, 6562-6579 (2013).

[24]   S. Krödel, T. Delpero, A. Bergamini, P. Ermanni, and D. M. Kochmann, *3D Auxetic Microlattices with Independently Controllable Acoustic Band Gaps and Quasi-Static Elastic Moduli*, Engineering Materials **16** (4), 357-363 (2014).

[25]   L. D'Alessandro, V. Zega, R. Ardito, and A. Corigliano, *3D Auxetic Single Material Periodic Structure with Ultra-Wide Tunable Bandgap*, Scientific Reports **8**, 2262 (2018).

[26]   D. Shin, Y. Urzhumov, D. Lim, K. Kim, and D. R. Smith, *A Versatile Smart Transformation Optics Device with Auxetic Elasto-Electromagnetic Metamaterials*, Scientific Reports **4**, 4084 (2014).

[27]   S. Barik, A. Karasahin, C. Flower, T. Cai, H. Miyake, W. DeGottardi, M. Hafezi, and E. Waks, *A Topological Quantum Optics Interface*, Science **359** (6376), 666-668 (2018).

[28]   M. I. Shalaev, S. Desnavi, W. Walasik, and N. M. Litchinitser, *Reconfigurable Topological Photonic Crystal*, New Journal of Physics **20**, 023040 (2018).

[29]   S. Mittal, J. Fan, S. Faez, A. Migdall, J. M. Taylor, and M. Hafezi, *Topologically Robust Transport of Photons in a Synthetic Gauge Field*, Physical Review Letters **113**, 087403 (2014).

[30]   J. Noh, W. A. Benalcazar, S. Huang, M. J. Collins, K. P. Chen, T. L. Hughes, and M. C. Rechtsman, *Topological Protection of Photonic Mid-Gap Defect Modes*, Nature Photonics **12**, 408–415 (2018).

[31]   M. L. N. Chen, L. J. Jiang, Z. Lan, and W. E. I. Sha, *Pseudospin-Polarized Topological Line Defects in Dielectric Photonic Crystals*, IEEE Transactions on Antennas and Propagation **68** (1), 609-613 (2020).

[32]   M. Z. Hasan and C. L. Kane, *Colloquium: Topological Insulators*, Reviews of Modern Physics **82**, 3045 (2010).

[33]   X. L. Qi and S. C. Zhang, *Topological Insulators and Superconductors*, Reviews of Modern Physics **83**, 1057 (2011).

[34]   L. Lu, J. D. Joannopoulos, and M. Soljačić, *Topological Photonics*, Nature Photonics **8** (11), 821-829 (2014).

[35]   S. D. Huber, *Topological Mechanics*, Nature Physics **12**, 621-623 (2016).

[36]   H. Huang, J. Chen, and S. Huo, *Recent Advances in Topological Elastic Metamaterials*, , Journal of Physics Condensed Matter **33**, 503002 (2021).

[37]   E. Prodan and C. Prodan, *Topological Phonon Modes and Their Role in Dynamic Instability of Microtubules*, Physical Review Letters **103**, 248101 (2009).

[38]   L. M. Nash, D. Kleckner, A. Read, V. Vitelli, A. M. Turner, and W. T. M. Irvine, *Topological Mechanics of Gyroscopic Metamaterials*, Proceedings of the National Academy of Sciences USA **112**, 14495–14500 (2015).

[39]   R. Fleury, A. B. Khanikaev, and A. Alù, *Floquet Topological Insulators for Sound*, Nature Communications **7**, 11744 (2016).

[40]   C. He, X. Ni, H. Ge, X. C. Sun, Y. Bin Chen, M. H. Lu, X. P. Liu, and Y. F. Chen, *Acoustic Topological Insulator and Robust One-Way Sound Transport*, Nature Physics **12**, 1124–1129 (2016).

[41]   M. Miniaci, R. K. Pal, B. Morvan, and M. Ruzzene, *Experimental Observation of Topologically Protected Helical Edge Modes in Patterned Elastic Plates*, Physical Review X **8**, 031074 (2018).

[42]   J. Lu, C. Qiu, L. Ye, X. Fan, M. Ke, F. Zhang, and Z. Liu, *Observation of Topological Valley Transport of Sound in Sonic Crystals*, Nature Physics **13**, 369–374 (2017).

[43]   R. Fleury, D. L. Sounas, C. F. Sieck, M. R. Haberman, and A. Alù, *Sound Isolation and Giant Linear Nonreciprocity in a Compact Acoustic Circulator*, Science **343** (6178), 516-519 (2014).

[44]   A. Souslov, B. C. van Zuiden, D. Bartolo, and V. Vitelli, *Topological Sound in Active-Liquid Metamaterials*, Nature Physics **13**, 1091–1094 (2017).

[45] C. Brendel, V. Peano, O. J. Painter, and F. Marquardt, *Pseudomagnetic Fields for Sound at the Nanoscale*, Proceedings of the National Academy of Sciences USA **114**, E3390-E3395 (2017).

[46] S. H. Mousavi, A. B. Khanikaev, and Z. Wang, *Topologically Protected Elastic Waves in Phononic Metamaterials*, Nature Communications **6**, 8682 (2015).

[47] R. K. Pal and M. Ruzzene, *Edge Waves in Plates with Resonators: An Elastic Analogue of the Quantum Valley Hall Effect*, New Journal of Physics **19**, 025001 (2017).

[48] R. K. Pal, M. Schaeffer, and M. Ruzzene, *Helical Edge States and Topological Phase Transitions in Phononic Systems Using Bi-Layered Lattices*, Journal of Applied Physics **119**, 084305 (2016).

[49] S. G. Haslinger, S. Frecentese, and G. Carta, *Localized Waves in Elastic Plates with Perturbed Honeycomb Arrays of Constraints*, Philosophical Transactions of the Royal Society A: Mathematical, Physical and Engineering Sciences **380**, 20210404 (2022).

[50] M. Miniaci and R. K. Pal, *Design of Topological Elastic Waveguides*, Journal of Applied Physics **130**, 141101 (2021).

[51] Y. Jin, D. Torrent, and B. Djafari-Rouhani, *Robustness of Conventional and Topologically Protected Edge States in Phononic Crystal Plates*, Physical Review B **98**, 054307 (2018).

[52] K. Sun, A. Souslov, X. Mao, and T. C. Lubensky, *Surface Phonons, Elastic Response, and Conformal Invariance in Twisted Kagome Lattices*, Proceedings of the National Academy of Sciences USA **109**, 12369-12374 (2012).

[53] C. L. Kane and T. C. Lubensky, *Topological Boundary Modes in Isostatic Lattices*, Nature Physics **10**, 39-45 (2013).

[54] J. Paulose, B. G. G. Chen, and V. Vitelli, *Topological Modes Bound to Dislocations in Mechanical Metamaterials*, Nature Physics **11**, 153-156 (2015).

[55] M. Miniaci, R. K. Pal, R. Manna, and M. Ruzzene, *Valley-Based Splitting of Topologically Protected Helical Waves in Elastic Plates*, Physical Review B **100**, 024304 (2019).

[56] Y. F. Wang, Y. Z. Wang, B. Wu, W. Chen, and Y. S. Wang, *Tunable and Active Phononic Crystals and Metamaterials*, Applied Mechanics Reviews **72** (4), 040801 (2020).

[57] N. K. Mohammadi, P. I. Galich, A. O. Krushynska, and S. Rudykh, *Soft Magnetoactive Laminates: Large Deformations, Transverse Elastic Waves and Band Gaps Tunability by a Magnetic Field*, Journal of Applied Mechanics, Transactions ASME **86** (11), 111001 (2019).

[58] A. S. Gliozzi, M. Miniaci, A. Chiappone, A. Bergamini, B. Morin, and E. Descrovi, *Tunable Photo-Responsive Elastic Metamaterials*, Nature Communications **11**, 2576 (2020).

[59] J. Hwan Oh, I. Kyu Lee, P. Sik Ma, and Y. Young Kim, *Active Wave-Guiding of Piezoelectric Phononic Crystals*, Applied Physics Letters **99**, 083505 (2011).

[60] B. I. Popa and S. A. Cummer, *Non-Reciprocal and Highly Nonlinear Active Acoustic Metamaterials*, Nature Communications **5**, 3398 (2014).

[61] A. Darabi, M. Collet, and M. J. Leamy, *Experimental Realization of a Reconfigurable Electroacoustic Topological Insulator*, Proceedings of the National Academy of Sciences USA **117**, 16138-16142 (2020).

[62] M. Gei, A. B. Movchan, and D. Bigoni, *Band-Gap Shift and Defect-Induced Annihilation in Prestressed Elastic Structures*, Journal of Appl Physics **105**, 063507 (2009).

[63] M. Miniaci, M. Mazzotti, A. Amendola, and F. Fraternali, *Effect of Prestress on Phononic Band Gaps Induced by Inertial Amplification*, International Journal of Solids and Structures **216**, 156-166 (2021).

[64] M. Mazzotti, I. Bartoli, and M. Miniaci, *Modeling Bloch Waves in Prestressed Phononic Crystal Plates*, Frontiers in Materials **6**, (2019).

[65] K. Bertoldi and M. C. Boyce, *Mechanically Triggered Transformations of Phononic Band Gaps in Periodic Elastomeric Structures*, Physical Review B **77**, 052105 (2008).

[66] J. R. Raney, N. Nadkarni, C. Daraio, D. M. Kochmann, J. A. Lewis, and K. Bertoldi, *Stable Propagation of Mechanical Signals in Soft Media Using Stored Elastic Energy*, Proceedings of the National Academy of Sciences USA **113**, 9722-9727 (2016).

[67] K. Bertoldi, V. Vitelli, J. Christensen, and M. van Hecke, *Flexible Mechanical Metamaterials*, Nature Reviews Materials **2**, 17066 (2017).

[68] M. Kheybari, C. Daraio, and O. R. Bilal, *Tunable Auxetic Metamaterials for Simultaneous Attenuation of Airborne Sound and Elastic Vibrations in All Directions*, Applied Physics Letters **121**, 081702 (2022).

[69] G. Carta, M. Brun, and A. Baldi, *Design of a Porous Material with Isotropic Negative Poisson's Ratio*, Mechanics of Materials **97**, 67-75 (2016).

[70] A. Baldi, M. Brun, and G. Carta, *Three-Dimensional Auxetic Porous Medium*, Mechanics of Materials **164**, 104114 (2022).

[71] A. Auld, *Acoustic Fields and Waves in Solids*, Vol. I (John Wiley and Sons, New York, 1973).

[72] E. N. Peters, *Thermoplastics, Thermosets, and Elastomers-Descriptions and Properties*, in *Mechanical Engineers' Handbook* (2015).

[73] S. Li, I. Kim, S. Iwamoto, J. Zang, and J. Yang, *Valley Anisotropy in Elastic Metamaterials*, Physical Review B **100**, 195102 (2019).

[74] P. Wang, L. Lu, and K. Bertoldi, *Topological Phononic Crystals with One-Way Elastic Edge Waves*, Physical Review Letters **115**, 104302 (2015).

[75] R. W. Ogden, *Non-Linear Elastic Deformations.* (1997).

[76] M. M. Attard, *Finite Strain - Isotropic Hyperelasticity*, International Journal of Solids and Structures **40** (17), 4353-4378 (2003).